\documentclass[a4paper]{spie}  

\usepackage{amsmath,amsfonts,amssymb}
\usepackage[colorlinks=true, allcolors=blue]{hyperref}

\RequirePackage{float}
\RequirePackage{graphicx, wrapfig}
\RequirePackage{caption}
\RequirePackage{subcaption}
\usepackage[percent]{overpic}
\usepackage{mhchem}

\title{Fast surrogate modelling of EIT in atomic quantum systems using LSTM neural networks}

\author[a]{Isabel S. Burdon Hita}
\author[a]{Óscar Iglesias-González}
\author[a]{Gabriel M. Carral}
\author[a]{Miguel Ferreira-Cao}
\affil[a]{Galician Research and Development Center in Advanced Telecommunications (GRADIANT) Carretera do Vilar, 56-58 36214 Vigo, Spain}

\authorinfo{Further author information:\\Isabel S. Burdon Hita: E-mail: iburdon@gradiant.org\\ Miguel Ferreira-Cao: E-mail: mferreira@gradiant.org}

\begin{document} 
\maketitle

\begin{abstract}
Simulations of optical quantum systems are essential for the development of quantum technologies. However, these simulations are often computationally intensive, especially when repeated evaluations are required for data fitting, parameter estimation, or real-time feedback. To address this challenge, we develop a Long Short-Term Memory neural network capable of replicating the output of these simulations with high accuracy and significantly reduced computational cost. Once trained, the surrogate model produces spectra in milliseconds, providing a speed-up of 5000x relative to traditional numerical solvers. We focus on applying this technique to Doppler-broadened Electromagnetically Induced Transparency in a ladder-type scheme for Rydberg-based sensing, achieving near-unity agreement with the physics solver for resonant and off-resonant regimes. We demonstrate the effectiveness of the LSTM model on this representative optical quantum system, establishing it as a surrogate tool capable of supporting real-time signal processing and feedback-based optimisation.

\end{abstract} 


\section{INTRODUCTION}
\label{sec:intro}
Successful engineering of quantum systems requires accurate modelling of the underlying physical principles. For instance, device prototyping and real-time data processing are dependent on precise yet efficient simulations. However, implementing accurate models is often a computationally demanding or intractable problem. Many quantum platforms rely on light-matter interactions that need to be described using density-matrix formalisms and, in particular, simulated by solving the Optical Bloch Equations (OBE)\cite{steck_quantum_2007}. This approach captures the dynamics of coherence, dissipation, and coupling to external fields with high accuracy, but it becomes computationally intensive when extended to realistic scenarios that incorporate inhomogeneous broadening, many-body effects, or multilevel atomic structures with complex coupling schemes. As a result, repeated evaluations required for parameter fitting, optimisation, or real-time feedback are often impractical when relying on direct numerical solvers. Such challenges are well illustrated in systems based on Electromagnetically Induced Transparency (EIT)\cite{fleischhauer_electromagnetically_2005,finkelstein_practical_2023}, especially when more than two atomic levels are involved. While high-performance computing resources could mitigate these issues, methods that run efficiently on standard, off-the-shelf hardware are still needed, reflecting the practical requirements of deployable quantum  technologies.

In this context, there has been growing academic interest in developing faster yet accurate approaches, including approximate methods and reduced models that aim to balance accuracy with computational efficiency\cite{xu_fast_2024}. Machine learning offers a means of efficiently replicating the full simulations without sacrificing accuracy\cite{gokhale_deep_2024, liu_deep_2022, wang_neural_2025}. By learning surrogate models of the system behaviour, machine learning can reproduce simulation results at a fraction of the computational cost, thereby allowing fast prediction of spectra, adaptive optimisation, and data analysis workflows that would otherwise be prohibitively slow. Recurrent neural networks, in particular those based on Long Short-Term Memory (LSTM) architectures\cite{hochreiter_long_1997}, are well suited to this task because they can efficiently capture correlations in structured data\cite{sahu_surrogate_2025, hajisharifi_lstm-enhanced_2024} such as spectra.

In this work, we present a machine learning framework for fast forward-modelling of atomic quantum systems. Although the methodology is general and can be applied to a broad class of platforms where the density matrix formalism is used, here we focus on the specific case of Rydberg-atom sensors for radio-frequency (RF) electrometry\cite{sedlacek_microwave_2012}. This system provides a useful test-bed for the approach, as accurate modelling of transmission spectra is critical for characterizing sensitivity and for supporting applications where real-time signal analysis is required. We show that the proposed LSTM surrogate achieves high agreement with numerical simulations while reducing evaluation time by more than three orders of magnitude. In addition, we show that the LSTM neural network generalizes well across both resonant and off-resonant regimes, extending its applications in experimental workflows.

\subsection{Rydberg Atomic Sensors}
Rydberg atoms have proved to be a powerful physical platform for quantum RF sensing\cite{artusio-glimpse_modern_2022}. Their strong dipole moments make them highly sensitive to external fields, providing direct measurements of RF electric fields without the need for calibration against classical antennas\cite{holloway_broadband_2014}. The working principle for Rydberg-atom sensors relies on EIT in a ladder-type scheme, where an applied RF field couples two Rydberg states and produces Autler–Townes (AT) splitting of the EIT resonance. The magnitude of the splitting is directly proportional to the RF Rabi frequency and therefore encodes the amplitude of the external field\cite{holloway_electric_2017}. This mechanism allows Rydberg sensors to operate as atom-based detectors that, relative to classical antennas, have certain performance advantages\cite{fancher_rydberg_2021, liu_electric_2023}, including wide tunability\cite{hu_continuously_2022} and high sensitivity\cite{simons_rydberg_2021, fan_atom_2015, meyer_assessment_2020}, without requiring substantial hardware modifications. These capabilities are realized with a common experimental platform: two counter-propagating laser beams interacting with a room-temperature alkali vapour. From this foundation, Rydberg sensors have been adapted to a range of applications, including radar imaging and ranging\cite{watterson_imaging_2025, chen_high-resolution_2025}, mm-wave chip testing\cite{borowka_rydberg-atom-based_2024}, and remote environmental monitoring such as soil moisture detection\cite{arumugam_remote_2024}.

The basic mechanism of Rydberg RF sensing relies on the coupling between highly-excited atomic states in alkali vapours. Their high dipole moments imply that even a small field can induce a strong coupling. In a ladder-type EIT scheme, a weak probe laser and a strong coupling laser, arranged as in Fig.~\ref{fig:Sensor_diagrams}(a), create a narrow transparency window. When an RF field illuminates the cell, it drives a transition between two Rydberg states. The EIT peak then splits into two distinct peaks through the Autler–Townes (AT) effect. The frequency separation $\Delta f$ of the AT peaks is directly proportional to the RF Rabi frequency $\Omega_{\mathrm{RF}}$ and hence to the field amplitude $E_{\mathrm{RF}}$\cite{holloway_electric_2017} :
\begin{equation}
\label{eq:AT_peak_relation}
\Delta f = \frac{\lambda_c}{\lambda_p}\frac{\Omega_{\mathrm{RF}}}{2\pi} \quad \text{with} \quad \Omega_{\mathrm{RF}} = \frac{d \, E_{\mathrm{RF}}}{\hbar} \, ,
\end{equation}
where $d$ is the transition dipole moment between the Rydberg states, while $\lambda_c$ and $\lambda_p$ are the control and probe wavelengths. The prefactor $\lambda_c/\lambda_p$ arises from the residual Doppler shift in a room-temperature vapour, reflecting the mismatch between the probe and control wavelengths in the two-photon resonance condition. Equation~\ref{eq:AT_peak_relation} provides a direct, atom-based measurement of the RF field strength, independent of calibration standards.

While this approach is straightforward, the extraction of RF parameters from spectra is computationally demanding when using full density-matrix simulations. To address this, machine learning offers a means of learning the nonlinear mapping between experimental parameters and simulation outputs\cite{cai_deep_2025}.

\subsection{Long Short-Term Memory Models}
In this work, we employ a LSTM neural network as a surrogate model. LSTMs are a class of recurrent network designed to capture long-range dependencies in sequential data\cite{hochreiter_long_1997}. Unlike convolutional or densely connected architectures, the LSTM can efficiently learn correlations between distant spectral regions.

Mathematically, a LSTM unit processes a sequence by maintaining two internal states: the hidden state $h_t$ and the cell state $c_t$, which together allow the network to preserve and update information across time steps. The key innovation of the LSTM is the use of gating mechanisms that regulate the flow of information: the input gate, the forget gate, and the output gate. At each time step $t$, given an input vector $x_t$ and the previous hidden and cell states $(h_{t-1}, c_{t-1})$, the LSTM updates its internal representation through the set of gates. Each gate applies a linear transformation, parametrized by a weight matrix $W_\ast$ and a bias vector $b_\ast$. These weights and biases are the trainable parameters of the network, learnt during optimisation.

In more detail, the forget gate $f_t$ determines which part of the past information is retained (Eq.~\ref{eq:forget_gate}), while the input gate $i_t$ (Eq.~\ref{eq:input_gate}) controls which new information is added through a candidate cell update $\tilde{c}_t$ (Eq.~\ref{eq:cell_activation}). The cell state is then updated according to Eq.~\ref{eq:updated_cell_state} so that memory from the past can be preserved while incorporating new information. Finally, the output gate $o_t$ (Eq.~\ref{eq:output_gate}) regulates how much of the updated cell state contributes to the hidden state (Eq.~\ref{eq:hidden_state}).
\begin{equation}
\label{eq:forget_gate}
    f_t = \sigma \left( W_f \cdot [h_{t-1}, x_t] + b_f \right) \, ,
\end{equation}
\begin{equation}
\label{eq:input_gate}
    i_t = \sigma \left( W_i \cdot [h_{t-1}, x_t] + b_i \right) \, , 
\end{equation}
\begin{equation}
\label{eq:cell_activation}
    \tilde{c}_t = \tanh \left( W_c \cdot [h_{t-1}, x_t] + b_c \right) \, ,
\end{equation}
\begin{equation}
\label{eq:updated_cell_state}
c_t = f_t \odot c_{t-1} + i_t \odot \tilde{c}_t \, ,
\end{equation}
\begin{equation}
\label{eq:output_gate}
o_t = \sigma \left( W_o \cdot [h_{t-1}, x_t] + b_o \right) \, ,
\end{equation}
\begin{equation}
\label{eq:hidden_state}
h_t = o_t \odot \tanh(c_t) \, .
\end{equation}
In the above equations, $\sigma(\cdot)$ denotes the logistic sigmoid, $\tanh(\cdot)$ is the hyperbolic tangent, and $\odot$ indicates element-wise multiplication.

The hidden state $h_t$ serves as the network’s output at each step, while the cell state $c_t$ functions as a persistent memory that evolves under the joint control of the gates. This architecture allows the LSTM to selectively preserve long-term dependencies while avoiding the vanishing or exploding gradient problem that affects conventional recurrent neural networks.

In the context of spectral modelling, the sequential structure of the LSTM allows the network to capture correlations across frequency points, ensuring that the predicted spectra remain smooth and physically consistent across the full range of detunings.

\section{METHODS}
\label{sec:methods}

\subsection{Physics-based Simulations}

\begin{figure}[H]
    \begin{minipage}[t]{.52\textwidth}
        \begin{overpic}[width=1\textwidth]{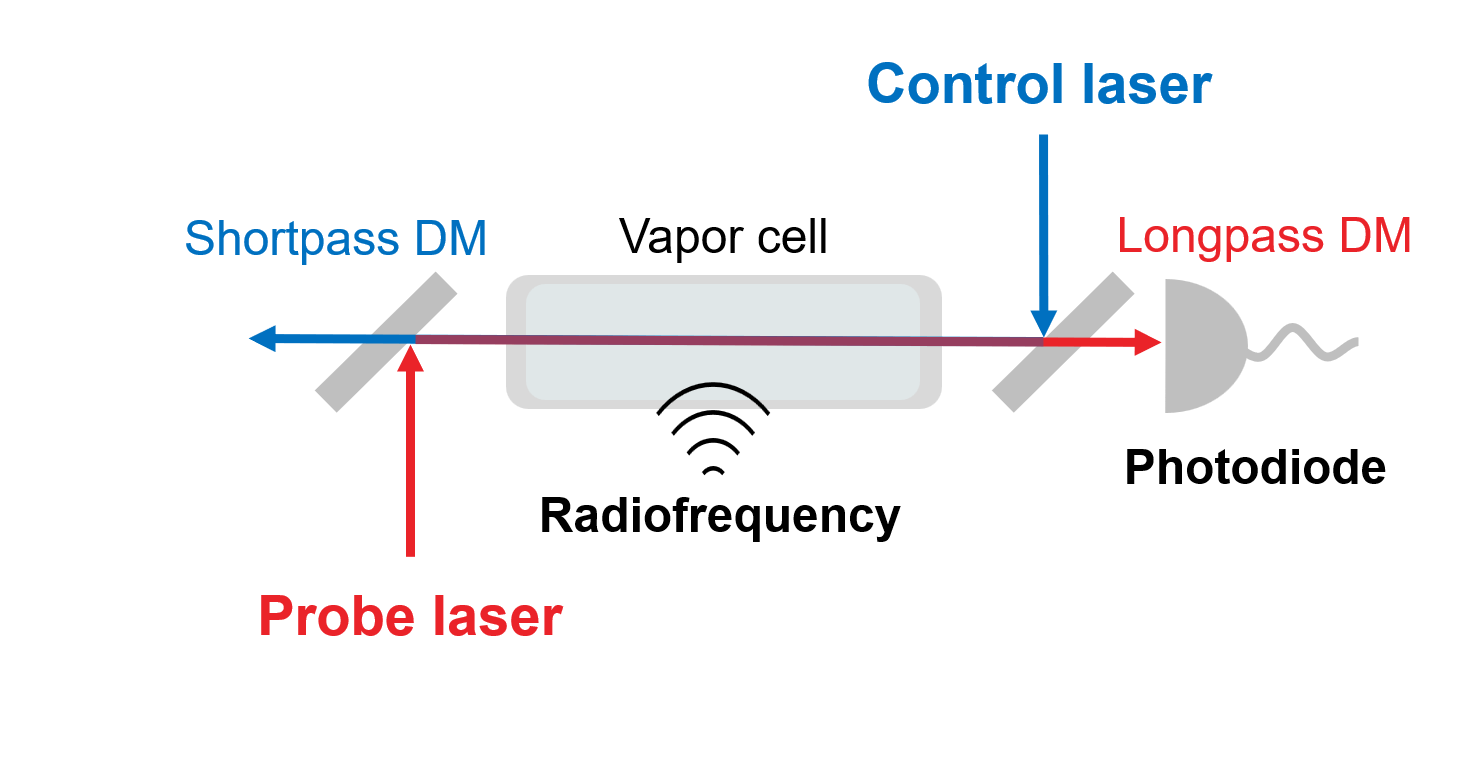}
        \put (2,2) {\small{(a)}}
        \end{overpic}
    \end{minipage}%
    \begin{minipage}[t]{.48\textwidth}
        \begin{overpic}[width=1
        \textwidth]{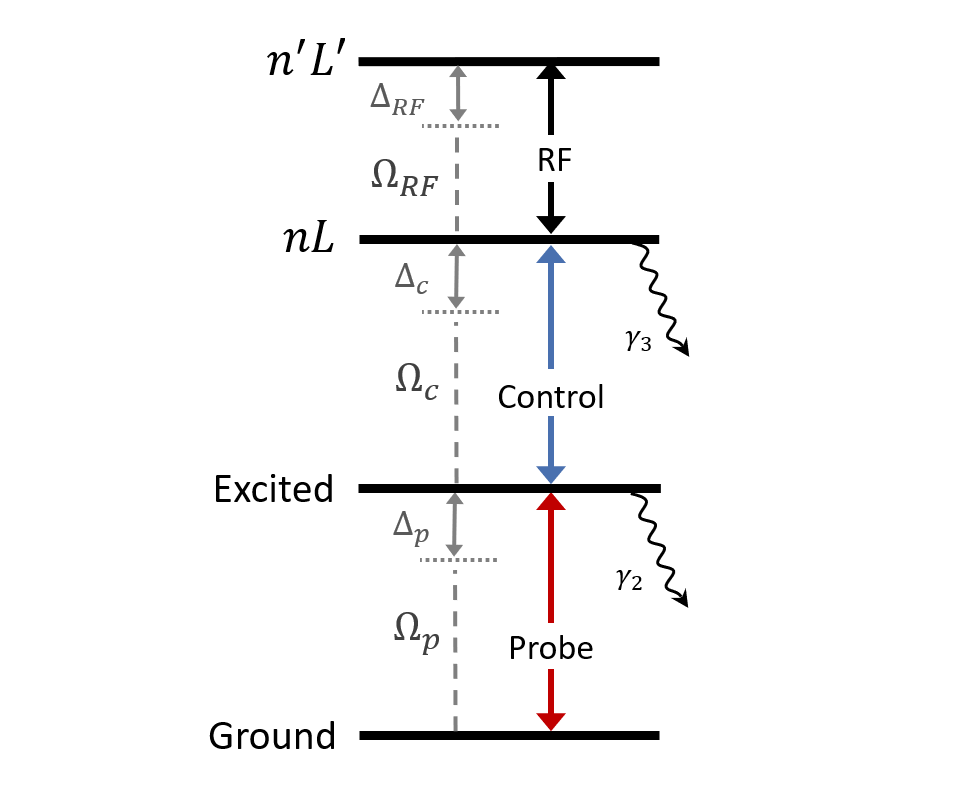}
        \put (2,2) {\small{(b)}}
        \end{overpic}
    \end{minipage}
    {\caption{(a) Basic experimental arrangement for a Rydberg sensor. Two counter-propagating and counter-aligned (overlapping) lasers meet at a vapour cell, reflected by the corresponding dichroic mirrors (DM). The cell is simultaneously illuminated by an external RF field, altering the EIT condition. The probe laser light is collected by a photodiode, thus providing an optical read-out of the RF strength according to the magnitude of the AT splitting. (b) Atomic levels of a typical alkali-metal Rydberg RF sensor. Ladder-type configuration with probe ($\Omega_p$), control ($\Omega_c$), and RF ($\Omega_{\mathrm{RF}}$) couplings. Key detunings, decoherence rates, and experimental parameters (atomic number density, cell length, temperature) are included in the simulation.} \label{fig:Sensor_diagrams}}
\end{figure}

We model the response of a three-level ladder-type Rydberg atom system coupled to an external RF signal, schematically shown in Fig.~\ref{fig:Sensor_diagrams}(b), by solving the optical Bloch equations\cite{steck_quantum_2007} in the steady state. The system dynamics are described by the Born-Markov master equation\cite{carmichael_statistical_1999} (Eq.~\ref{eq:master_equation}):
\begin{equation}
\label{eq:master_equation}
    \dot{\rho} = -\frac{i}{\hbar}[H,\rho] + \mathcal{L}(\rho) \, ,
\end{equation}
\begin{equation}
\label{eq:Lindbladian}
    \mathcal{L}(\rho) = \sum_{i,j}\frac{\Gamma_{ij}}{2}\left(2\sigma_{ij}\rho\sigma_{ij}^\dagger - \sigma_{ij}^\dagger\sigma_{ij}\rho - \rho\sigma_{ij}^\dagger\sigma_{ij}\right) \, ,
\end{equation}
where $H$ is the system's Hamiltonian in the rotating-wave approximation, $\mathcal{L}(\rho)$ accounts for spontaneous decay and dephasing processes, $\Gamma_{ij}$ are the decay rates and $\sigma_{ij}$ are the set of jump operators. For the ladder configuration considered here, the Hamiltonian includes the probe, coupling, and RF interactions with corresponding Rabi frequencies $\Omega_p$, $\Omega_c$, and $\Omega_{\mathrm{RF}}$, plus the detunings $\Delta_p$, $\Delta_c$ and $\Delta_{\mathrm{RF}}$. The explicit form of $H$\cite{holloway_electric_2017} is given in Eq.~\ref{eq:hamiltonian}:

\begin{equation}
\label{eq:hamiltonian}
H = \frac{\hbar}{2}
    \begin{pmatrix}
    0 & \Omega_p & 0 & 0\\
    \Omega_p & -2\Delta_p & \Omega_c & 0\\
    0 & \Omega_c & -2(\Delta_p + \Delta_c) & \Omega_{\mathrm{RF}}\\
    0 & 0 & \Omega_{\mathrm{RF}} & -2(\Delta_p + \Delta_c + \Delta_{\mathrm{RF}})\\
    \end{pmatrix} \, .
\end{equation}

From the master equation, we obtain the coupled OBEs for the density matrix elements. These equations amount to a set of 16 coupled complex equations for the four-level system.

For each parameter configuration, the steady-state solution of the OBEs is obtained numerically via LU (lower-upper) decomposition. To account for Doppler effects, the Maxwell–Boltzmann velocity distribution is discretized into 850 velocity classes, efficiently calculated using the inverse probability distribution, covering probabilities down to $10^{-3}$. For each velocity, the detunings in the Hamiltonian are redefined to include the corresponding Doppler shift, the OBEs recalculated, and the steady-state solution obtained. The probe susceptibility is then computed from the coherence between ground and intermediate states, and the transmittance\cite{steck_quantum_2007} through the atomic vapour cell is finally given by
\begin{equation}
\label{eq:transmittance}
|T|^2 = e^{- k_p L \, \mathrm{Im}[\chi]} \, ,
\end{equation}
where $k_p = 2\pi/\lambda_p$ is the probe wavenumber, $L$ is the cell length, and $\chi$ is the Doppler-averaged probe susceptibility. A full transmission spectrum with 300 frequency points therefore requires $850 \times 300$ evaluations. Using a fully optimised and vectorized Python implementation, the runtime on a standard CPU\footnote{Simulations were performed on an Intel Core i7-1260P (12 cores, 16 threads, base frequency 2.1 GHz), without GPU acceleration.} is on the order of 5–10 minutes per spectrum.

\begin{figure}[H]
    \begin{minipage}[t]{.5\textwidth}
        \begin{overpic}[width=1\textwidth]{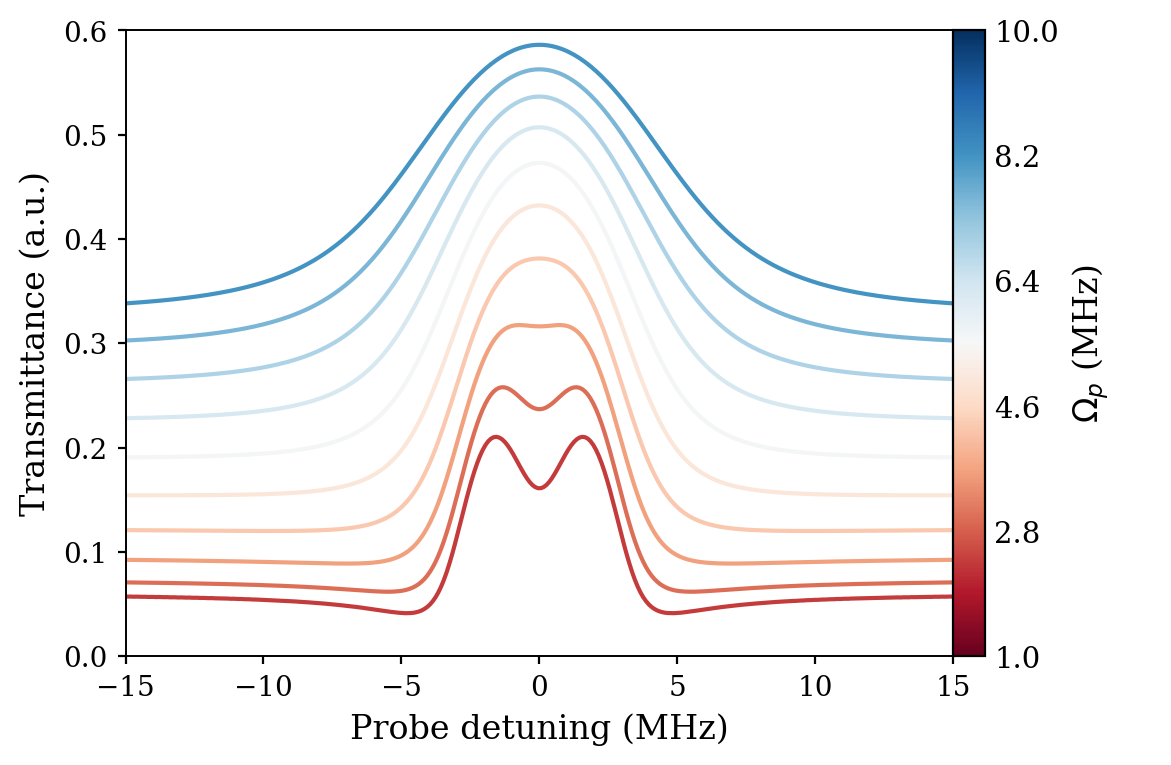}
        \put (2,2) {\small{(a)}}
        \end{overpic}
    \end{minipage}%
    \begin{minipage}[t]{.5\textwidth}
        \begin{overpic}[width=1\textwidth]{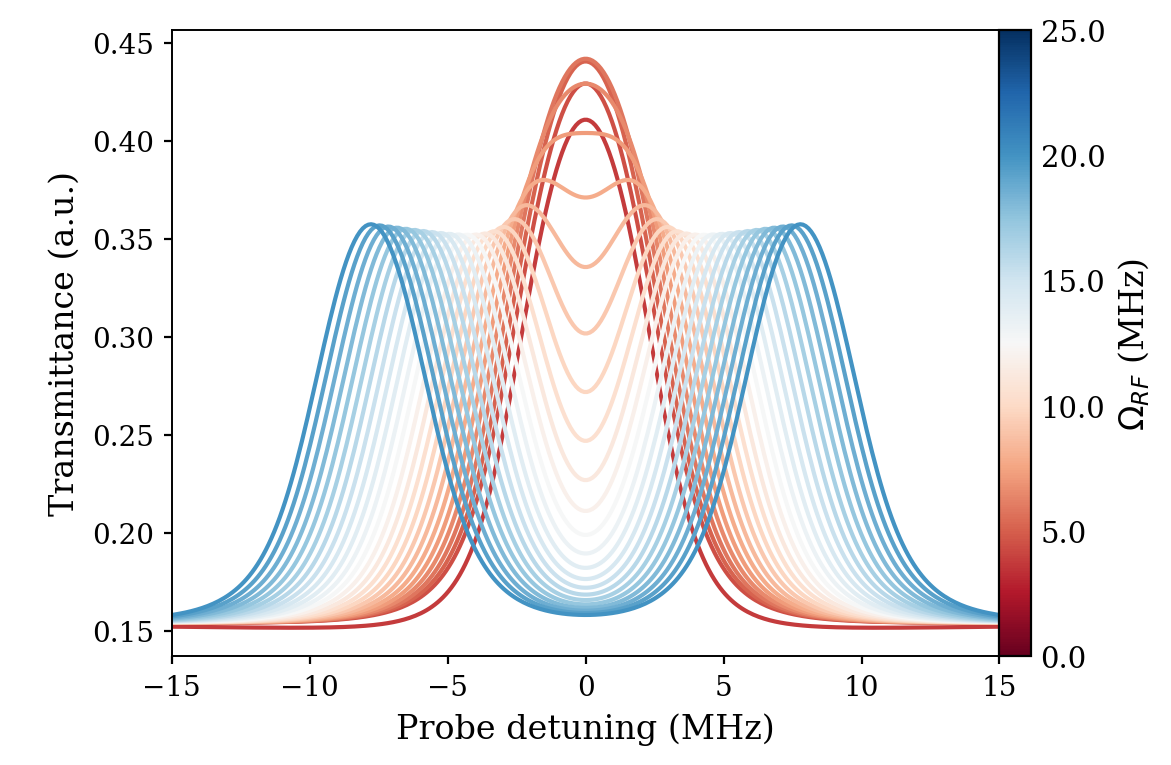}
        \put (2,2) {\small{(b)}}
        \end{overpic}
    \end{minipage}
    {\caption{Representative spectra from the physics-based solver using $\ce{^{87}Rb}$ atomic parameters\cite{steck_rubidium_2003, sibalic_arc_2017}. Control Rabi frequency: 6 MHz. Temperature: 288 K. Cell length: 10 cm. (a) Variation with probe Rabi frequency, illustrating saturation broadening effects. (b) Variation with RF Rabi frequency, showing Autler–Townes splitting.} \label{fig:dataset_sweeps}}
\end{figure}

To train the surrogate model, we generated a synthetic dataset of 250 simulated spectra, each with 300 frequency points, following the methodology described above. The parameters swept in the dataset were the probe Rabi frequency (1–10 MHz\footnote{All Rabi frequencies are expressed in angular units (e.g., $1\,\text{MHz} \equiv 2\pi \times 10^6\,s^{-1}$)}) and the RF Rabi frequency (0–25 MHz). Representative spectra illustrating the dependence on probe and RF amplitudes are shown in (Fig.~\ref{fig:dataset_sweeps}). Both inputs and outputs were normalized to comparable magnitudes prior to training.

\subsection{Surrogate Model}
\begin{figure}[h]
    \centering
    \includegraphics[width=13cm]{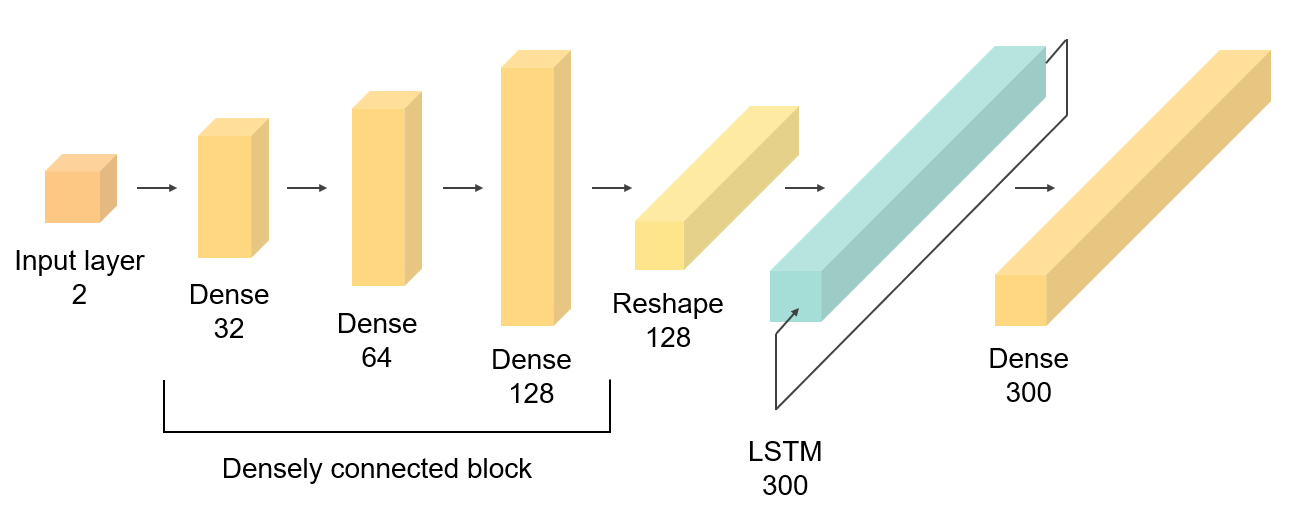}
    {\caption{Architecture of the surrogate LSTM model. Inputs are probe and RF Rabi frequencies. The architecture includes three dense layers (32, 64, 128 nodes), an LSTM block (300 outputs), and a final dense layer (300 linear outputs). Total model size is $\sim$7 MB.} \label{fig:surrogate_model_architecture}}
\end{figure}

The surrogate model is implemented as a compact neural network with an architecture centred on a LSTM layer\cite{hochreiter_long_1997} (Fig.~\ref{fig:surrogate_model_architecture}). The two physical parameters, the probe Rabi frequency $\Omega_p$ and the RF Rabi frequency $\Omega_{\mathrm{RF}}$, serve as inputs, while the network output is the full probe transmission spectrum with 300 points, matching the frequency grid of the physics solver.

In contrast to conventional architectures, where recurrent layers act as encoders of high-dimensional input sequences\cite{lipton_critical_2015}, here the LSTM layer serves directly as the generative core of the model. The physical parameters are first embedded through a set of densely connected layers, producing a compact representation of the experimental conditions. This representation then drives the LSTM, which sequentially unfolds the spectral output across probe detuning. The recurrence captures both local spectral features, such as line shape and linewidth, and long-range structure, such as baseline shifts and Autler–Townes splitting.

By inverting the typical recurrent neural network structure, this architecture is both lightweight and well adapted to the physics problem: the low-dimensional parameter space is mapped onto a high-dimensional but strongly correlated output. The LSTM is therefore not used to compress sequential data, but rather to generate spectra that are consistent with the underlying optical response. This repurposing of recurrent networks highlights a novel way for constructing surrogate models of quantum-optical systems.

Specifically, the network architecture consists of an initial encoding block with three densely connected layers (comprised of 32, 64, and 128 nodes), followed by a single LSTM layer with 300 units, and a final dense layer with 300 outputs to produce the spectrum (Fig.~\ref{fig:surrogate_model_architecture}). The activations for the layers are linear, with the exception of the first two layers in the encoding block which use ReLU activations for efficiency. The total model size is approximately 7 MB, making it lightweight to train and straightforward to deploy.

Training used the Huber loss function, which combines the advantages of mean squared error (MSE) for small deviations and mean absolute error (MAE) for larger deviations\cite{gokcesu_generalized_2021}. For an error term $a = y - \hat{y}$, the Huber loss is defined as
\begin{equation}
\label{eq:Huber_loss}
L_\delta(a) =
\begin{cases}
\frac{1}{2} a^2, & |a| \leq \delta \\
\delta \left(|a| - \frac{1}{2}\delta\right), & |a| > \delta
\end{cases} \, .
\end{equation}
This loss function is particularly well-suited for spectral data, where both local accuracy and robustness to large variations are important.

The dataset was divided 85/15 into training and validation sets, with additional spectra reserved for independent testing. Using 250 spectra proved sufficient to achieve excellent predictive accuracy. This highlights the training efficiency of the network in capturing spectral features and its capacity to generalize reliably, despite the relatively small size of the training data by machine-learning standards.

\section{RESULTS}
\label{sec:results}

The surrogate model achieves high predictive accuracy in the training and validation domains. Quantitatively, we obtain coefficients of determination $R^2 > 0.999$ and root mean squared error (RMSE) below 0.002 for the validation set, with inference times under 50 ms. The metrics used are defined as
\begin{equation}
R^2 = 1 - \frac{\sum_i (y_i - \hat{y}_i)^2}{\sum_i (y_i - \bar{y})^2} \, ,
\end{equation}
\begin{equation}
\text{RMSE} = \sqrt{\frac{1}{N} \sum_i (y_i - \hat{y}_i)^2} \, ,
\end{equation}
where $y_i$ are the reference values, $\hat{y}_i$ the predictions, and $\bar{y}$ their mean. Beyond these numerical scores, the model reproduces the main physical features of the spectra, including peak positions, widths, separations, and offsets, while maintaining smooth spectral profiles. This ensures that the surrogate can not only interpolate within the training set but also generalize reliably to unseen parameter regimes.

\subsection{Testing on Unseen Parameter Sets \& Global Performance}
\begin{figure}[h]
    \begin{minipage}[c]{.5\textwidth}
        \begin{overpic}[width=1\textwidth]{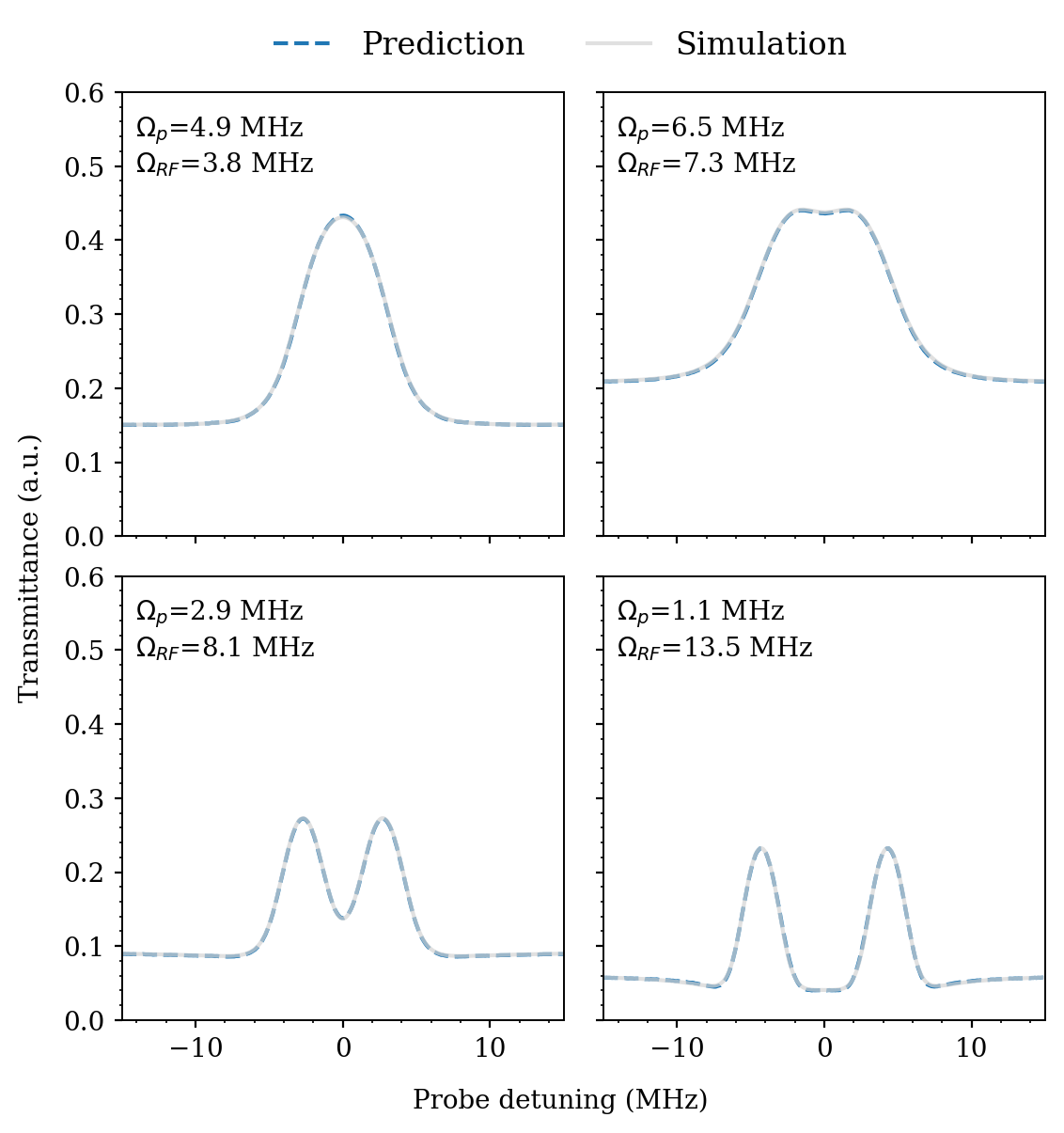}
        \put (2,2) {\small{(a)}}
        \end{overpic}
    \end{minipage}%
    \begin{minipage}[c]{.5\textwidth}
        \begin{overpic}[width=1\textwidth]{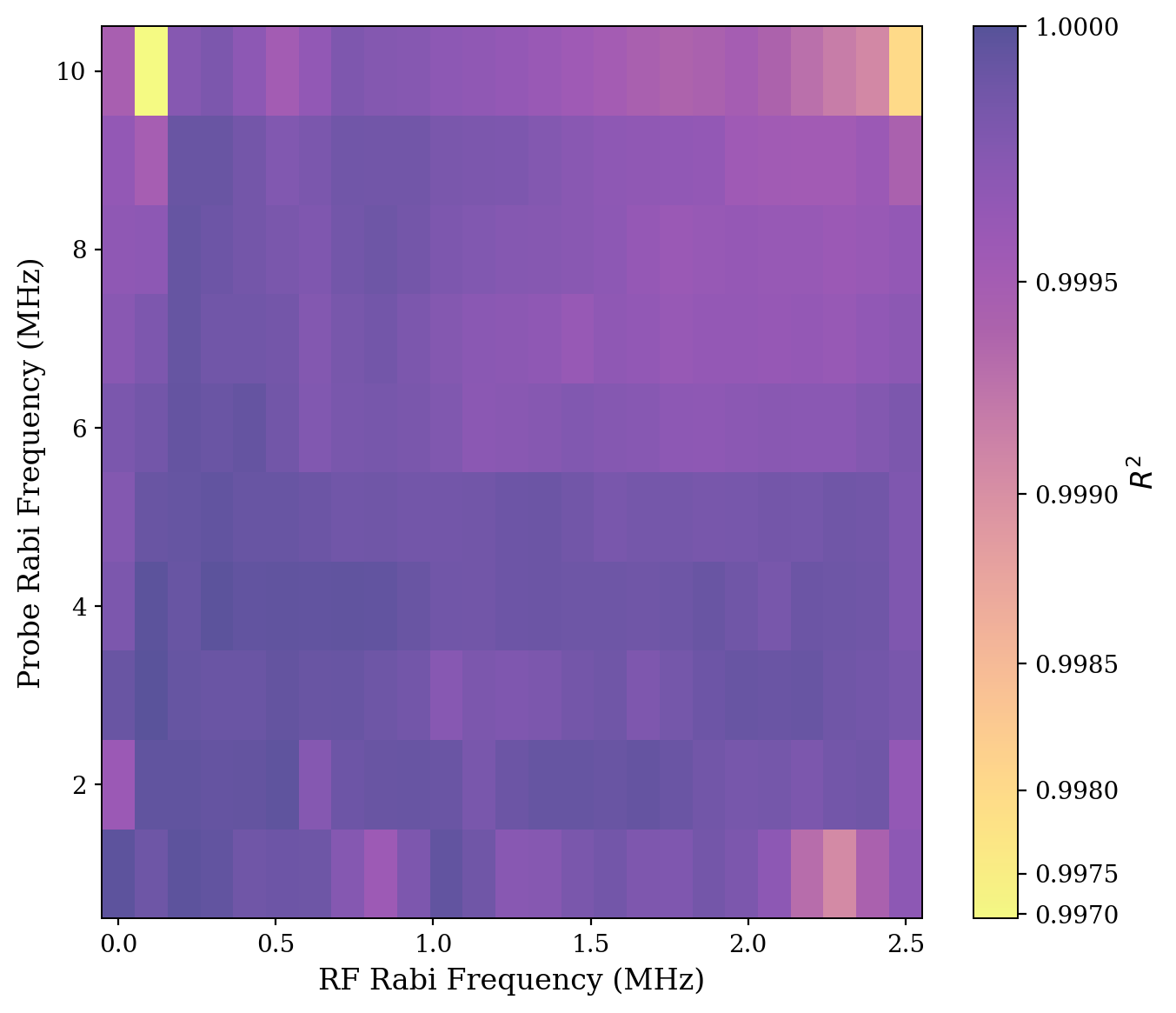}
        \put (2,-8.7) {\small{(b)}}
        \end{overpic}
    \end{minipage}
    {\caption{Results of surrogate model tests with $\ce{^{87}Rb}$ atomic parameters\cite{steck_rubidium_2003, sibalic_arc_2017}. The fixed parameters used were: $\Omega_c = 6$ MHz and $T=288$ K with a cell length of 10 cm. (a) Testing on four of the parameter sets not included in the training or validation data. Solid lines: physics solver; dashed lines: surrogate predictions. The surrogate successfully captures amplitude variations, peak widths, separations, and offsets, while preserving smooth spectral shapes. (b) Heatmap of $R^2$ performance across the parameter grid. The surrogate model achieves $R^2>0.999$ across most of the domain, with a slight performance drop at the corners due to reduced sampling density. The model produces physically consistent outputs across the entire space.} \label{fig:resonance_results}}
\end{figure}

To evaluate the surrogate model beyond the training and validation set, we selected 10 representative test cases not included in the initial database of 250 spectra. To illustrate the results, the comparison between the physics-based solver (solid lines) and the surrogate predictions (dashed lines) for four of the samples is shown in Figure~\ref{fig:resonance_results}(a). The metrics obtained for the test set are in line with the validation set results, achieving an average RMSE under 0.001 and an average $R^2>0.999$.

There is excellent agreement between the two models across all testing points. The surrogate model not only captures the main spectral features, such as peak positions and relative amplitudes, but also reproduces fine variations in amplitude, peak width, peak separation, and overall transmittance offset. Importantly, the surrogate model preserves the smoothness of the spectral curves, which is essential for subsequent post-processing tasks such as fitting and peak finding. Achieving this level of detail is non-trivial: non-recurrent architectures struggle to generate long sequences with stable correlations across output nodes. By contrast, the recurrent nature of the LSTM allows the surrogate to produce coherent outputs over the full 300-point spectrum. These results demonstrate that the surrogate model is able to generalize well and retain physical consistency, even for cases outside its training dataset.

To further quantify the model's accuracy, we evaluated the surrogate on the entire dataset of generated spectra, comprising both training and validation points. Figure \ref{fig:resonance_results}(b) shows the resulting heatmap of the coefficient of determination ($R^2$) across the parameter space. The surrogate achieves near-unity scores over most of the domain, with $R^2>0.999$ in the central regions.

A slight drop in performance is visible toward the corners of the parameter grid. This reduction is expected: at the edges of the parameter space, fewer neighbouring points are available for interpolation during training, making predictions more challenging. Nevertheless, the surrogate continues to produce physically consistent spectra even in these regions, without spurious artifacts or divergences. This suggests that, while denser sampling would further improve edge performance, the current database already provides robust coverage for most practical regimes of interest.

\subsection{LSTM Adequacy for Complex Physical Regimes: Off-resonant RF Sensing}
Beyond the resonant Autler–Townes regime, where peak splitting scales linearly with the RF amplitude, Rydberg sensors can also be operated under off-resonant driving conditions. This regime is particularly attractive for continuous frequency sensing\cite{simons_continuous_2021} and can, in principle, enhance sensitivity by exploiting the increased separation between the AT peaks in off-resonance\cite{simons_using_2016}. In practice, however, quantitative analysis in this regime is challenging: the spectra exhibit pronounced asymmetries, shifts in peak positions, and detuning-dependent changes in peak separation, meaning that the simple proportionality between peak distance and RF Rabi frequency no longer holds\cite{zhang_detuning_2019}. As a result, conventional analysis techniques fail to extract RF amplitude and detuning information directly from a single measured spectrum and need to rely instead on impractical extrapolation techniques\cite{simons_using_2016}.

\begin{figure}[h]
    \begin{minipage}[c]{.5\textwidth}
        \begin{overpic}[width=1\textwidth]{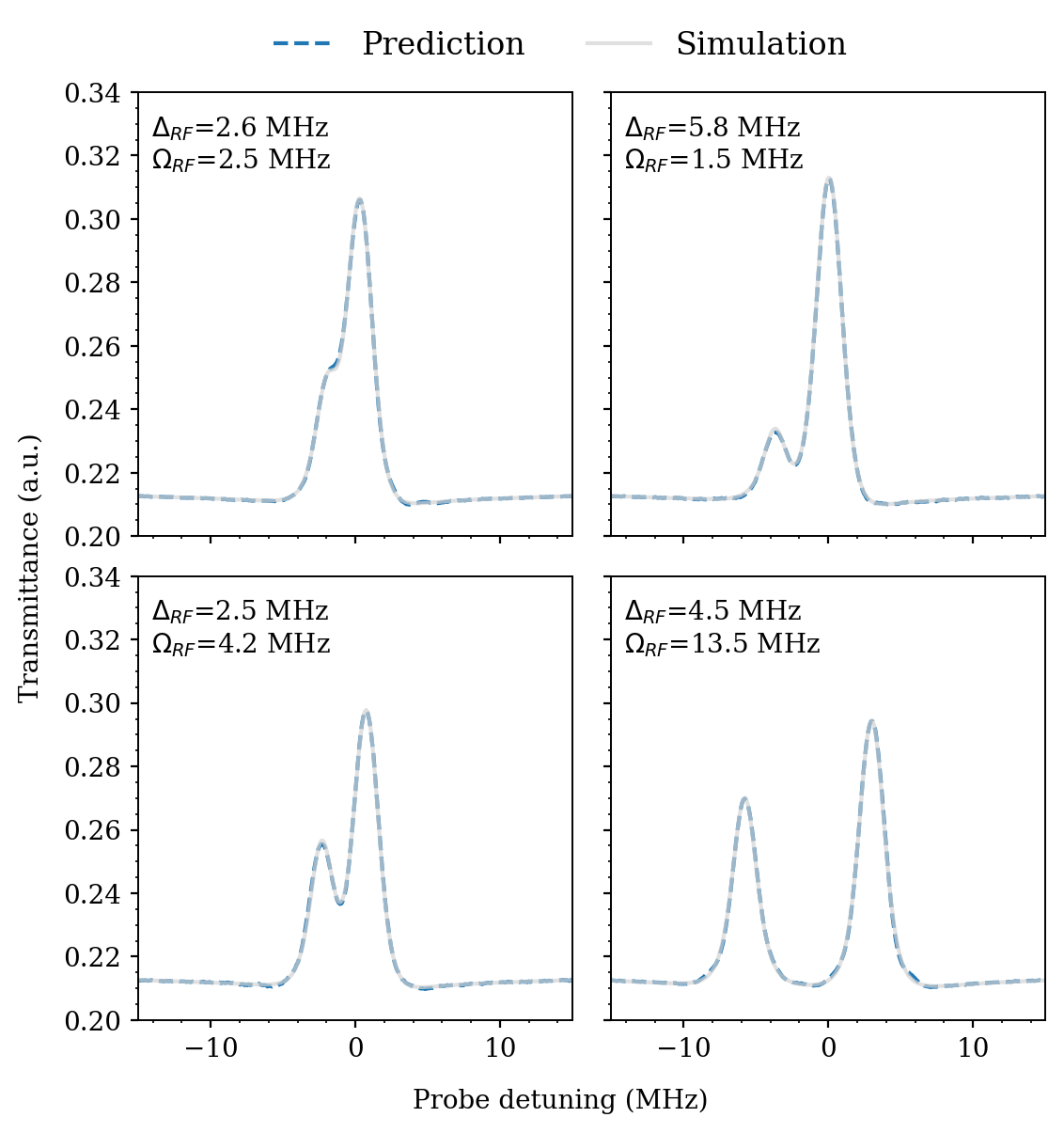}
        \put (2,2) {\small{(a)}}
        \end{overpic}
    \end{minipage}%
    \begin{minipage}[c]{.5\textwidth}
        \begin{overpic}[width=1\textwidth]{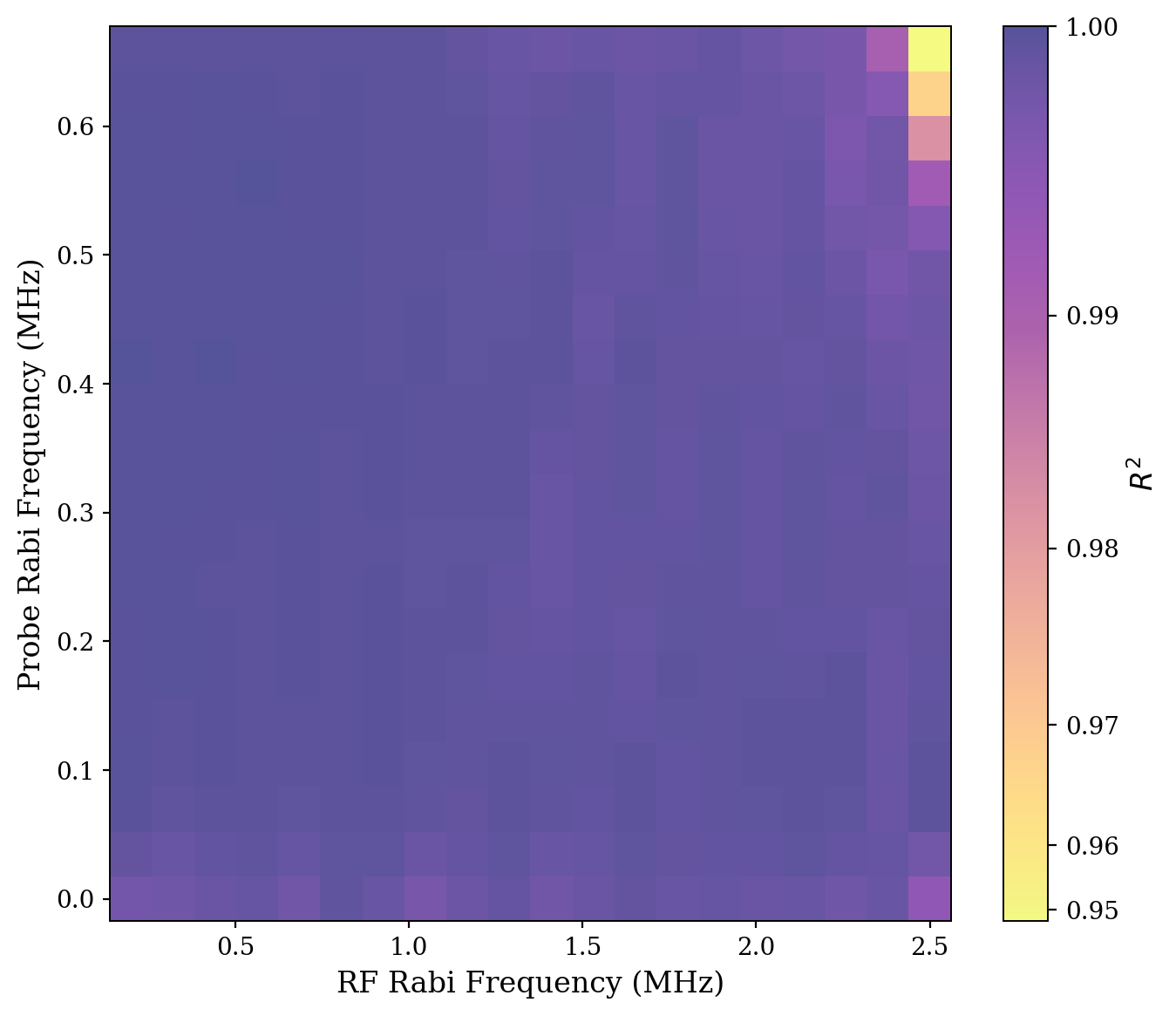}
        \put (2,-8.7) {\small{(b)}}
        \end{overpic}
    \end{minipage}
    {\caption{Off-resonant regime comparison using \textsuperscript{87}Rb atomic parameters\cite{steck_rubidium_2003, sibalic_arc_2017}. For the training and validation sets, the fixed parameters used were: $\Omega_c = 2$ MHz, $\Omega_p=1.5$ MHz and $T=283$ K with a cell length of 10 cm. (a) Example spectra comparing solver outputs (solid) and surrogate predictions (dashed). (b) Heatmap of $R^2$ performance across RF Rabi frequency and detuning. The surrogate generalizes well, maintaining high accuracy despite the increased spectral variance introduced by detuning.} \label{fig:off_resonance_results}}
\end{figure}

To explore this regime, we trained the surrogate on an extended dataset of 530 spectra spanning both the RF Rabi frequency and the magnitude of RF detuning. Figure~\ref{fig:off_resonance_results} shows representative comparisons between the solver and surrogate outputs, along with a heatmap of $R^2$ scores across the full parameter domain. Despite the substantially higher spectral variance, the surrogate accurately reproduces both global features and local variations, with the average $R^2>0.99$ and RMSE below 0.01 in both the test and validation sets.

These results demonstrate two key points. First, the LSTM surrogate can generalize beyond the relatively simple resonant case to describe more complex, multi-parameter regimes without modification of the network architecture. Second, the surrogate provides a new analysis pathway for off-resonant spectra, that is, it is a tool for parameter estimation in a regime where no direct analytical relations exist for measurable features in a single spectra. This opens possibilities for real-time frequency-resolved sensing and for optimising sensor operation in regimes that were previously impractical to work in due to the lack of efficient data analysis tools.

\subsection{Computational Efficiency}
Finally, we benchmarked the computational performance of the surrogate model against the physics solver. On a standard CPU, solving the full set of optical Bloch equations for each spectrum requires several minutes, depending on the number of velocity classes and detuning points included. In contrast, the trained surrogate model generates the same spectrum in $48.24 \pm 2.26$ ms, corresponding to a speedup in the order of 5000×.

This reduction in runtime is particularly significant for real-time applications, such as adaptive experiment optimisation, online fitting of experimental data, or closed-loop control in quantum sensing scenarios. In these contexts, the ability to obtain high-fidelity spectra within milliseconds transforms the feasibility of advanced control strategies that would otherwise be impossible with conventional numerical solvers.

\section{CONCLUSIONS AND OUTLOOK}
\label{sec:conclusions}
We have developed and evaluated a neural network surrogate model for Rydberg RF sensing, trained directly on spectra generated by a full steady-state solution of the optical Bloch equations with Doppler broadening. The surrogate model achieves excellent agreement with the physics solver across both resonant and off-resonant regimes, reproducing detailed spectral features while reducing evaluation times to milliseconds. In practice, the model provides real-time analysis of Rydberg sensor spectra on modest CPU hardware. By demonstrating that accurate spectral prediction can be achieved with standard processors, this work points toward deployable quantum technologies where automated sensors extract maximal information in situ. This efficiency is particularly relevant for portable or resource-constrained platforms\cite{liu_electric_2023, xing_chip-scale_2025, ma_mems_2025}, such as mobile, aerial, or satellite-based systems.

Several natural extensions arise from this work. For instance, training the surrogate model directly on experimental data could bypass the need for detailed physics modelling while still capturing system-specific features and providing insights on noise characterization. Applications of the surrogate model in the off-resonant regime may allow direct extraction of RF parameters from (single) measured spectra, a task that is not accessible with current analysis techniques. Integrating the surrogate into closed-loop control workflows could provide real-time optimisation of sensor sensitivity and dynamic adaptation to experimental drift. Results here can be adapted in order to include alternative Rydberg sensor architectures which exploit different signatures of the external RF on the optical read-out, such as non-linear\cite{sedlacek_microwave_2012} or superheterodyne detection\cite{jing_atomic_2020}.

Beyond Rydberg sensors, the same methodology can be extended to other quantum-optical systems where accurate but computationally expensive models currently limit real-time analysis. More broadly, machine-learning surrogates could accelerate not only the engineering of quantum devices but also their use as probes of fundamental physics. Potential applications include quantum state tomography\cite{james_measurement_2001,gavryusev_density_2016}, characterization of state preparation in quantum simulation \cite{ferreira-cao_depletion_2020}, and signal processing with quantum sensors, where real-time feedback could assist Hamiltonian estimation and system optimisation\cite{santagati_magnetic-field_2019}. Similar strategies could also support numerical studies of analog quantum simulators\cite{orioli_relaxation_2018, signoles_glassy_2021, schultzen_glassy_2022, franz_observation_2024}, where repeated evaluations with quantum and semi-classical numerical models are computationally intensive.

In summary, this work establishes a general methodology for employing machine-learning surrogates to accelerate the simulation of quantum systems, with immediate applications in quantum sensing, for real time signal processing of experimental data, and with potential extensions across a wider range of quantum technologies.

\acknowledgments      
This research was supported by the Spanish Centre for the Development of Industrial Technology (CDTI) and the Ministry of Economy, Industry and Competitiveness within Spanish ``Plan de Recuperación, Transformación y Resiliencia", co-financed by the European Union - NextGeneration UE, under grant/project CER-20231018 ``6G Distributed and Federated Experiments in Radio, Edge and Non-Terrestrial Networks" (6GDIFERENTE).

\bibliography{report} 
\bibliographystyle{spiebib} 

\end{document}